\documentclass[3p,times,twocolumn]{elsarticle}

\usepackage{ecrc}

\volume{00}

\firstpage{1}

\journalname{Nuclear Physics B Proceedings Supplement}

\runauth{}

\jid{nppp}

\jnltitlelogo{Nuclear Physics B Proceedings Supplement}

\usepackage{amssymb}
\usepackage{bm}

\usepackage{lineno}

\biboptions{comma,sort&compress}

\usepackage[figuresright]{rotating}

\begin{document}

\begin{frontmatter}



\dochead{}

\title{Exact solutions of the Boltzmann equation and optimized hydrodynamic approaches for relativistic heavy-ion collisions\tnoteref{tn0}}
\tnotetext[tn1]{Work supported by the U.S. Department of Energy, Office of Science, Office of Nuclear Physics (UH, DB, MM, and MS), by NSERC (GSD), and by CNPq and FAPESP (JN).}


\author[a]{U.~Heinz\fnref{fn1}}
\fntext[fn1]{Presenter. E-mail: heinz.9@osu.edu.}
\author[a]{D.~Bazow}
\author[b]{G.~S.~Denicol}
\author[a]{M.~Martinez}
\author[c]{M.~Nopoush}
\author[d,e]{J.~Noronha}
\author[f]{R.~Ryblewski}
\author[c]{M.~Strickland}
\address[a]{Physics Department, The Ohio State University, Columbus, OH 43210-1117, USA}
\address[b]{Department of Physics, McGill University, 3600 University Street, Montreal,
QC, H3A 2T8, Canada}
\address[c]{Department of Physics, Kent State University, Kent, OH 44242 United States}
\address[d]{Instituto de F\'{\i}sica, Universidade de S\~{a}o Paulo, C.P. 66318, 05315-970 S\~{a}o Paulo, SP, Brazil}
\address[e]{Department of Physics, Columbia University, 538 West 120th Street, New York, NY 10027, USA}
\address[f]{The H. Niewodnicza\'nski Institute of Nuclear Physics,
Polish Academy of Sciences, PL-31342 Krak\'ow, Poland}

\begin{abstract}
Several recent results are reported from work aiming to improve the quantitative precision of relativistic viscous fluid dynamics for relativistic heavy-ion collisions. The dense matter created in such collisions expands in a highly anisotropic manner. Due to viscous effects this also renders the local momentum distribution anisotropic. Optimized hydrodynamic approaches account for these anisotropies already at leading order in a gradient expansion. Recently discovered exact solutions of the relativistic Boltzmann equation in anisotropically expanding systems provide a powerful testbed for such improved hydrodynamic approximations. We present the latest status of our quest for a formulation of relativistic viscous fluid dynamics that is optimized for applications to relativistic heavy-ion collisions. 
\end{abstract}

\begin{keyword}
quark-gluon plasma \sep viscous fluid dynamics \sep Boltzmann equation \sep relativistic heavy-ion collisions


\end{keyword}

\end{frontmatter}


\section{Motivation}
\label{sec1}

Relativistic viscous hydrodynamics has become the workhorse of dynamical modeling of ultra-relativistic heavy-ion collisions. It is an effective macroscopic description based on coarse-graining (via a gradient expansion) of the underlying microscopic dynamics. Its systematic construction is still a matter of debate, complicated by the existence of a complex hierarchy of micro- and macroscopic time scales that are not well separated in relativistic heavy-ion collisions. Exact solutions of the highly nonlinear microscopic dynamics can serve as a testbed for macroscopic hydrodynamic approximation schemes. Such solutions have been found for the Boltzmann equation in the Relaxation Time Approximation (RTA), which describes weakly interacting systems, under the assumption of highly symmetric flow patterns and density distributions (Bjorken \cite{Bjorken:1982qr} and Gubser \cite{Gubser:2010ze
} flows) of the system \cite{Baym:1984np,Florkowski:2013lza,
Florkowski:2014sfa,Denicol:2014xca,
Heinz:2015cda}. Here we use both to test different hydrodynamic expansion schemes.

\vspace*{-3mm}
\section{Kinetic theory vs. hydrodynamics}
\label{sec2}
\vspace*{-1mm}

Hydrodynamics is an effective theory whose form is independent of the microscopic interaction strength. Its equations can thus be derived from kinetic theory in a window of weak coupling and small pressure gradients where both approaches are simultaneously valid. [We will here use the RTA Boltzmann equation as our starting point.] Only the values of the transport coefficients and the equation of state depend on the microscopic coupling strength;  for the strongly coupled quark-gluon plasma created in heavy-ion collisions, they must be obtained with non-perturbative methods.

The conserved macroscopic currents $j^\mu=\langle p^\mu\rangle$ (particle current) and $T^{\mu\nu}=\langle p^\mu p^\nu\rangle$ (energy-momentum tensor) are obtained by taking momentum moments $\langle h(p)\rangle\equiv\frac{g}{(2\pi)^3}\int\frac{d^3p}{E_p} h(p) f(x,p)$ of the distribution function $f(x,p)$ ($g$ is the degeneracy factor.) Hydrodynamic equations are obtained by splitting the distribution function into a leading-order contribution $f_0$, parametrized through macroscopic observables as \cite{Martinez:2012tu,Tinti:2013vba,Nopoush:2015yga}
\begin{equation}
\label{eq1}
f_0(x,p)=f_0\left(\frac{\sqrt{p_\mu\Xi^{\mu\nu}(x)p_\nu}-\tilde\mu(x)}{\tilde{T}(x)}\right),
\end{equation}
and a smaller first-order correction $\delta f$ ($|\delta f/f_0|\ll1$):
\begin{equation}
\label{eq2}
f(x,p)=f_0(x,p)+\delta f(x,p).
\end{equation}
In Eq.~(\ref{eq1}), $\Xi^{\mu\nu}(x)=u^\mu(x) u^\nu(x)-\Phi(x)\Delta^{\mu\nu}(x)+\xi^{\mu\nu}(x)$, where the hydrodynamic flow field $u^\mu(x)$ defines the local fluid rest frame (LRF) and $\Delta^{\mu\nu}=g^{\mu\nu}{-}u^\mu u^\nu$ is the spatial projector in the LRF. $\Phi(x)$ and the tensor $\xi(x)$ partially account for bulk viscous effects and shear-viscous deviations from local momentum isotropy in anisotropically expanding systems. $\tilde{T}(x)$, $\tilde{\mu}(x)$ are the effective temperature and chemical potential in the LRF. 

$u^\mu(x)$, $\tilde{T}(x)$, and $\tilde{\mu}(x)$ are fixed by the Landau matching conditions, requiring $u^\mu u_\mu=1$:
\begin{equation}
\label{eq3}
T^\mu_{\ \nu} u^\nu = \mathcal{E} u^\mu,\quad
\langle u{\cdot}p\rangle_{\delta f} = \langle (u{\cdot}p)^2\rangle_{\delta f} = 0.
\end{equation}
Here the eigenvalue $\mathcal{E}(\tilde{T},\tilde{\mu};\xi,\Phi)$ is the LRF energy density. The true local temperature $T(\tilde{T},\tilde{\mu};\xi,\Phi)$ and local chemical potential $\mu(\tilde{T},\tilde{\mu};\xi,\Phi)$ are introduced by demanding $\mathcal{E}(\tilde{T},\tilde{\mu};\xi,\Phi)=\mathcal{E}_\mathrm{eq}(T,\mu)$ and $\mathcal{N}(\tilde{T},\tilde{\mu};\xi,\Phi)\equiv\langle u{\cdot}p\rangle_{f_0}=\mathcal{R}_0(\xi,\Phi)\mathcal{N}_\mathrm{eq}(T,\mu)$ where $\mathcal{E}_\mathrm{eq}$, $\mathcal{N}_\mathrm{eq}$ are the thermal equilibrium energy and particle densities and $\mathcal{R}_0$ is a factor that depends on the viscous deformations $\xi$ and $\Phi$ of the local momentum distribution \cite{Martinez:2010sc,Tinti:2013vba,Nopoush:2015yga}.

Writing $T^{\mu\nu}= T^{\mu\nu}_0+\delta T^{\mu\nu}\equiv T^{\mu\nu}_0+\Pi^{\mu\nu}$, $j^\mu= j^\mu_0+\delta j^\mu\equiv j^\mu_0+V^\mu$, the conservation laws
\begin{equation}
\label{eq4}
\partial_\mu T^{\mu\nu}(x)=0,\quad 
\partial_\mu j^\mu(x)=\frac{{\cal N}(x)-{\cal N}_\mathrm{eq}(x)}{\tau_\mathrm{rel}(x)}
\end{equation}
are sufficient to determine $u^\mu(x)$, $T(x)$, $\mu(x)$, but not the dissipative corrections $\xi^{\mu\nu},\,\Phi,\,\Pi^{\mu\nu},$ and $V^\mu$; their evolution is controlled by microscopic physics. Different hydrodynamic approaches can be distinguished by the assumptions they make about the dissipative corrections and/or the approximations they use to derive their dynamics from the underlying Boltzmann equation:\\[0.2ex]
{\bf 1.\ Ideal hydrodynamics} assumes local momentum isotropy of $f$, setting $f_0$ to be isotropic ($\xi^{\mu\nu}=0$) and all dissipative currents to zero: $\Phi=\Pi^{\mu\nu}=V^\mu=0$.\\[0.2ex]
{\bf 2.\ Navier-Stokes (NS) theory} maintains local momentum isotropy at leading order (i.e. in $f_0$), sets $\Phi{\,=\,}0$, and postulates instantaneous constituent relations for $\Pi^{\mu\nu}$ and $V^\mu$ by introducing viscosity and heat conduction as transport coefficients that relate these flows to their driving forces, but ignores the microscopic relaxation time that is needed for these flows to adjust to their Navier-Stokes values. This leads to acausal signal propagation.\\[0.2ex]
{\bf 3.\ Israel-Stewart (IS) theory} \cite{Israel:1979wp} improves on NS theory by evolving  $\Pi^{\mu\nu}$ and $V^\mu$ dynamically, with evolution equations derived from moments of the Boltzmann equation, keeping only terms linear in the Knudsen number $\mathrm{Kn}=\lambda_\mathrm{mfp}/\lambda_\mathrm{macro}$.\\[0.2ex]
{\bf 4.\ Denicol-Niemi-Molnar-Rischke (DNMR) theory} \cite{Denicol:2012cn} improves IS theory by keeping nonlinear terms up to order $\mathrm{Kn}^2,\, \mathrm{Kn}\cdot\mathrm{Re}^{-1}$ when evolving $\Pi^{\mu\nu}$ and $V^\mu$. (Terms of second order in the inverse Reynolds number $\mathrm{Re}^{-1}$ vanish in the RTA used here but would otherwise appear, too.)\\[0.2ex]
{\bf 5.\ Anisotropic hydrodynamics ({\sc aHydro})} \cite{Martinez:2010sc
} allows allows for a leading-order local momentum anisotropy ($\xi^{\mu\nu},\,\Phi\ne0$), evolved according to equations obtained from low-order moments of the Boltzmann equation, but ignores residual dissipative flows: $\Pi^{\mu\nu}=V^\mu=0$.\\[0.2ex] 
{\bf 6.\ Viscous anisotropic hydrodynamics ({\sc vaHydro})} \cite{Bazow:2013ifa
} improves on {\sc aHydro} by additionally evolving (using IS or DNMR theory) the residual dissipative flows $\Pi^{\mu\nu},\,V^\mu$ generated by the deviation $\delta f$ around the locally anisotropic leading-order distribution function $f_0$.

\vspace*{-3mm}
\section{Exact solutions of the Boltzmann equation}
\label{sec3}
\vspace*{-2mm}
\subsection{Systems undergoing Bjorken flow}
\label{sec3a}

For highly symmetric flow profiles the Boltzmann equation can be solved exactly in RTA. Bjorken flow \cite{Bjorken:1982qr} describes the dynamics of a longitudinally boost invariant, transversally homogeneous system. A system with these symmetries is most conveniently discussed in Milne coordinates $(\tau,r,\phi,\eta)$ where $\tau=(t^2{-}z^2)^{1/2}$ and $\eta=\frac{1}{2}\ln[(t{-}z)/(t{+}z)]$. Bjorken flow is static in these coordinates, $u^\mu=(1,0,0,0)$. In Cartesian coordinates this implies a longitudinal flow velocity profile $v_z=z/t$ \cite{Bjorken:1982qr}. The metric in Milne coordinates is $ds^2=d\tau^2{-}dr^2-r^2d\phi^2-\tau^2d\eta^2$. The Bjorken symmetry restricts the possible dependence of the distribution function $f(x,p)$ to $f(x,p) = f(\tau;p_\perp,w)$ where $w=tp_z-zE=\tau m_\perp\sinh(y{-}\eta)$ is boost invariant. The RTA Boltzmann equation then simplifies to an ordinary differential equation
\begin{equation}
\label{eq6}
\partial_\tau f(\tau;p_\perp,w) = -\frac{f(\tau;p_\perp,w)-f_\mathrm{eq}(\tau;p_\perp,w)}
                                                           {\tau_\mathrm{rel}(\tau)}
\end{equation}
with the solution \cite{Baym:1984np,Florkowski:2013lza}                                                        
\begin{eqnarray}
\label{eq7}
\!\!\!\!\!\!
f(\tau;p_\perp,w)&=& D(\tau,\tau_0)f_0(p_\perp,w)
\\\nonumber
&+&\int_{\tau_0}^\tau\,\frac{d\tau'}{\tau_{\mathrm{rel}}(\tau')}\,
D(\tau,\tau')\,f_{\mathrm{eq}}(\tau';p_\perp,w)\quad
\end{eqnarray}
where $D(\tau_2,\tau_1)=\exp\!\left(-\int_{\tau_1}^{\tau_2}\,\frac{d\tau''}{\tau_{\mathrm{rel}}(\tau'')}\right)$. 

\subsection{Systems undergoing Gubser flow}
\label{sec3b}

For systems with longitudinal boost invariance and azimuthally symmetric radial dependence, Gubser \cite{Gubser:2010ze} found a flow pattern describing simultaneous longitudinal and transverse radial expansion by starting from a static system, $u^\mu=(1,0,0,0)$, in de Sitter coordinates $(\rho,\theta,\phi,\eta)$, with
\begin{eqnarray}
\label{eq8}
&&\rho(\tau,r) =-\sinh^{-1}\left( \frac{1-q^2\tau^2+q^2r^2}{2q\tau }\right),
\nonumber\\
&&\theta(\tau,r) =\tan^{-1}\left(\frac{2qr}{1+q^2\tau^2-q^2r^2}\right).
\end{eqnarray}
In Cartesian coordinates this leads to the flow profile $v_z=z/t$, $v_r= \frac{2q^{2}\tau r}{1{+}q^{2}\tau ^{2}{+}q^{2}r^{2}}$ where $q$ is an arbitrary scale parameter. The de Sitter metric is $d\hat{s}^2 \equiv ds^2/\tau^2 = d\rho^2{-}\cosh^2\!\rho\,(d\theta^{2}+\sin^{2}\theta\, d\phi^{2}) -d\eta^{2}$. A distribution function that preserves the symmetries of this metric can depend on $(x,p)$ only as $f(x,p) = f(\rho;\hat{p}_\Omega^2,\hat{p}_\eta)$ where $\hat{p}_\Omega^2=\hat{p}_\theta^2 + \frac{\hat{p}_\phi^2}{\sin^2\theta}$ and $\hat{p}_\eta=w$. Due to conformal symmetry of the metric and flow profile \cite{Gubser:2010ze} the particles described by $f$ must be massless. With $T(\tau,r)\equiv\hat{T}(\rho(\tau,r))/\tau$ the RTA Boltzmann equation simplifies to the ordinary differential equation 
\begin{eqnarray}
\label{eq8a}
\!\!\!\!\!\!
\frac{\partial}{\partial \rho} f(\rho;\hat{p}_\Omega^2,\hat{p}_\varsigma)
= -\frac{\hat{T}(\rho)}{c}\left[ f\!
\left(\rho;\hat{p}_\Omega^2,\hat{p}_\varsigma\right)
- f_{\mathrm{eq}}\!\left(\frac{\hat{p}^\rho}{\hat{T}(\rho)}\right) %
\right],
\nonumber
\end{eqnarray}
where for the conformal case $\tau_\mathrm{rel}(x)=c/T(x)$ and $c/5=\eta/s$ is the specific shear viscosity. The solution is \cite{Denicol:2014xca}
\begin{eqnarray}
\label{eq9}
f(\rho;\hat{p}^2_\Omega,w)&{=}& D(\rho,\rho_0)\, f_0(\hat{p}_\Omega^2,w) 
\\\nonumber
\quad
+\frac{1}{c}\int_{\rho_0}^\rho &{}&
\!\!\!\!\!\!\!\!\!\!\!\!\!\!\! 
d\rho'\,\hat{T}(\rho')\, D(\rho,\rho')\,f_\mathrm{eq}(\rho';\hat{p}_\Omega^2,w).\qquad
\end{eqnarray}
This integral equation in $\rho$ can be solved separately for each pair of momenta $(\hat{p}_\Omega^2,w)$. It can be worked out for any ``initial'' condition $f_0(\hat{p}_\Omega^2,w)\equiv f(\rho_0;\hat{p}_\Omega^2,w)$. We here use equilibrium initial conditions, $f_0=f_\mathrm{eq}$.

\subsection{Hydrodynamics of Gubser flow}
\label{sec3c}

By taking hydrodynamic moments, the exact $f$ yields the exact evolution of all components of $T^{\mu\nu}$. Here, due to the high degree of symmetry, $\Pi^{\mu\nu}$ has only one independent component, the longitudinal shear stress $\pi^{\eta\eta}$. This exact evolution, which reflects the exact solution of the microscopic dynamics, can be compared to solutions, in Milne or de Sitter coordinates, of the various hydrodynamic approximation discussed at the end of Sec.~\ref{sec2}, using identical initial conditions. For Bjorken flow, the equations corresponding to the hydrodynamic approximations {\bf 1.}--{\bf 6.} are given and their solutions compared to the exact result in \cite{Florkowski:2013lza,Florkowski:2014sfa,Bazow:2013ifa}. For Gubser flow one finds the following \cite{Denicol:2014xca,Marrochio:2013wla,Nopoush:2014qba}:\\[0.5ex]
{\bf 1.\ Ideal hydrodynamics:} $\hat{T}_{\mathrm{ideal}}(\rho) = \hat{T}_0/\cosh^{2/3}(\rho)$.\\[0.5ex]
{\bf 2.\ NS theory} solves instead the differential equ\-ation $(1/\hat{T})(d\hat{T}/d\rho)+\frac{2}{3}\tanh \rho = \frac{1}{3}\bar{\pi}_{\eta}^{\eta}(\rho )\,\tanh \rho$ for viscous $T$-evolution \cite{Gubser:2010ze,Marrochio:2013wla}, with $\bar{\pi}_{\eta}^{\eta} \equiv \hat{\pi}_\eta^\eta/(\hat{T}\hat{s})$ and $\hat{\pi}_{NS}^{\eta\eta}=\frac{4}{3}\hat{\eta} \,\tanh \rho$ where $\hat{\eta}/\hat{s}{\,=\,}\frac{1}{5}\hat{T}\hat{\tau}_\mathrm{rel}$ (all variables with hats are made unitless by multiplying with appropriate powers of $\tau$).\\[0.5ex]
{\bf 3.\ In IS theory} the instantaneous constituent equation for $\bar{\pi}_{\eta}^{\eta}$ in NS theory is replaced by a dynamical evolution equation \cite{Marrochio:2013wla}:\\[0.5ex]
 $d\bar{\pi}_{\eta}^{\eta}/d\rho + \frac{4}{3}\left( \bar{\pi}_{\eta}^{\eta}\right)^{2}\tanh \rho +\bar{\pi}_\eta^\eta/\hat{\tau}_\mathrm{rel} =\frac{4}{15}\tanh\rho$.\\[1ex]
{\bf 4.\ DNMR theory} adds to this evolution equation a term of order $\mathrm{Kn}{\cdot}\mathrm{Re}^{-1}$ \cite{Denicol:2014xca}:\\[0.5ex]
$d\bar{\pi}_{\eta}^{\eta}/d\rho + \frac{4}{3}\left(\bar{\pi}_{\eta}^{\eta}\right)^{2}\tanh\rho + \bar{\pi}_\eta^\eta/\hat{\tau}_\mathrm{rel} = \frac{4}{15} \tanh \rho 
+ \frac{10}{21}\bar{\pi}_\eta^\eta \tanh\rho$.\\[1ex]
{\bf 5.\ Anisotropic hydrodynamics ({\sc aHydro})} solves a different set of evolution equations for the temperature and pressure anisotropy which can be found in \cite{Nopoush:2014qba}.\\[0.5ex]
{\bf 6.\ For viscous anisotropic hydrodynamics ({\sc vaHydro})} the form of the hydrodynamic equations corresponding to Gubser symmetry have not yet been worked out.  

\vspace*{-3mm}
\section{Results}
\label{sec4}
\vspace*{-1mm}

\begin{figure}
\hspace*{-6mm}
\includegraphics[bb=0 0 550 420, width=1.1\linewidth,clip=,angle=0]{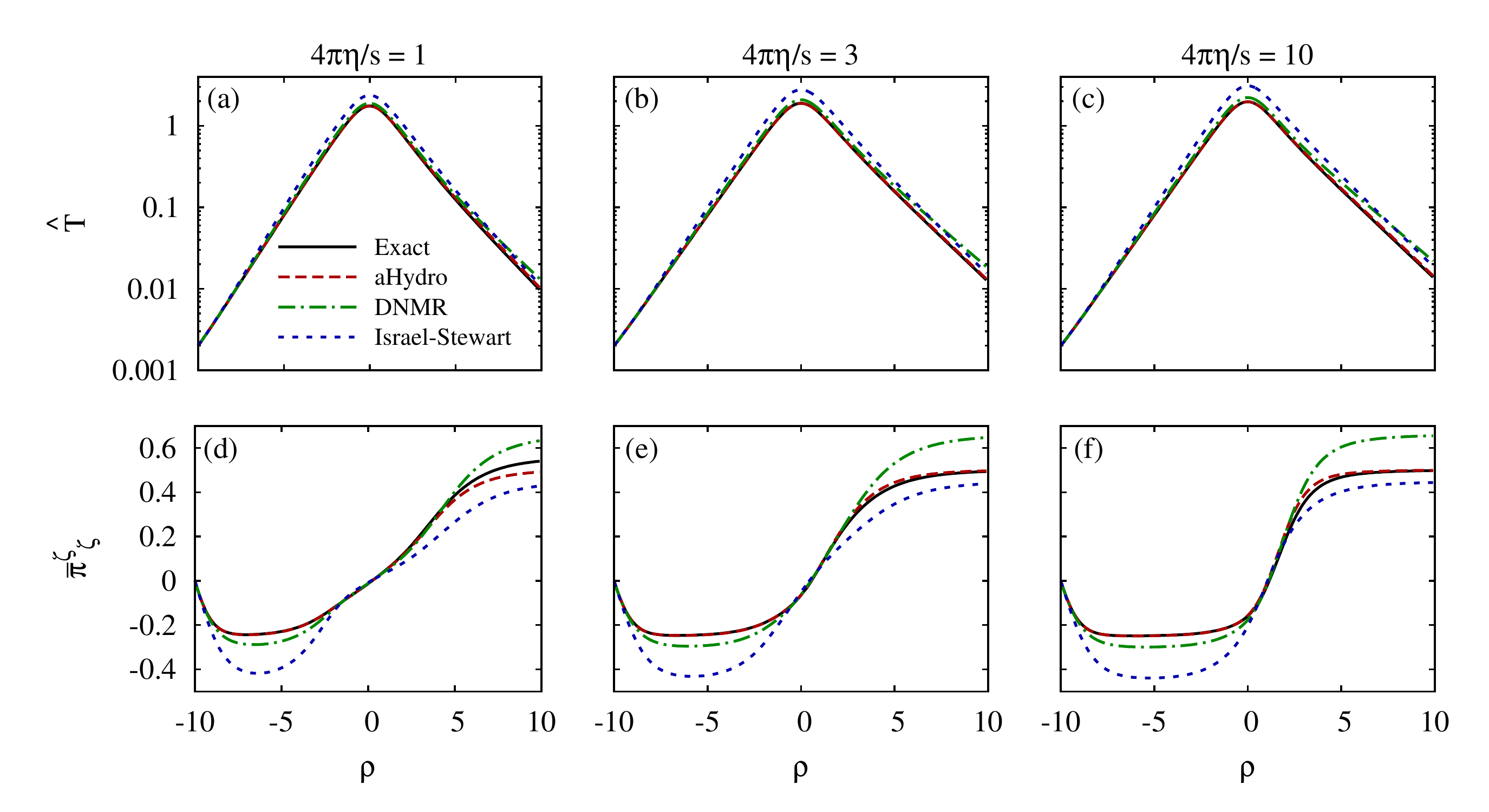}
\vspace*{-8mm}
\caption{de Sitter time ($\rho$) evolution of the effective temperature $\hat{T}$ (top row) and the scaled shear stress component $\bar{\pi}_{\eta}^{\eta}$ (denoted by $\bar{\pi}_{\varsigma}^{\varsigma}$ in the figure) (bottom row), for the exact solution of the Boltzmann equation (sold black line), Israel-Stewart (short-dashed blue line) and DNMR (dash-dotted green line) second-order viscous hydrodynamics, and anisotropic hydrodynamics (long-dashed red line). The two columns correspond to different specific shear viscosity values as indicated. Locally isotropic thermal equilibrium initial conditions with $\hat{T}=0.002$ were imposed at $\rho_0=-10$. (Taken from Ref.~\cite{Nopoush:2014qba}.)
\label{F1}}
\end{figure}

For Bjorken flow the comparison of the exact evolution with the approximations {\bf 1.}--{\bf 6.} was reported in \cite{Florkowski:2013lza,Florkowski:2014sfa,Bazow:2013ifa} and will not be repeated here due to space restrictions. It was found that NS theory reproduced the exact result better than ideal fluid dynamics, second-order viscous hydrodynamics (IS or DNMR theory) works better than first-order NS theory but failed badly for systems with very large viscosity (it leads to large entropy production even in the free-streaming, infinite viscosity limit), that anisotropic hydrodynamics repaired this deficiency and provided qualitatively correct descriptions of the exact result for both small and large viscosities, but that all these approaches were outperformed by {\sc vaHydro} which, for Bjorken flow, reproduces the exact result almost perfectly over the entire range of viscosities and initial conditions \cite{Bazow:2013ifa}.

Here we discuss one example from the analogous comparisons made in \cite{Denicol:2014xca,Nopoush:2014qba} for systems expanding with the Gubser flow profile. The comparison, shown in Fig.~\ref{F1}, is made in de Sitter coordinates -- the translation to Minkowski coordinates is straightforward, and corresponding plots can be found in \cite{Denicol:2014xca,Heinz:2015cda}. Similar to what was reported above for Bjorken flow, anisotropic hydrodynamics reproduces the exact microscopic Boltzmann dynamics much more accurately than standard second-order viscous hydrodynamics, especially for systems with large viscosity and pressure anisotropy (see \cite{Nopoush:2014qba}). This happens even before the effects from residual dissipative flows, resulting from the deviation of the distribution function from the leading-order ansatz $f_0$ made in {\sc aHydro}, are included. {\sc vaHydro} includes the latter but has not yet been worked out for Gubser flow. As in the Bjorken case, it is expected to further improve agreement with the exact solution.    

A recent analysis of the exact solution of the Boltzmann equation with Gubser symmetry showed that, for thermal equilibrium initial conditions at $\rho_0$, the solution becomes unphysical for large negative $\rho{-}\rho_0$, as a result of the distribution function turning negative for large transverse and small longitudinal momenta \cite{Heinz:2015cda}. With such initial conditions it should therefore only be trusted for positive de Sitter times, $\rho{-}\rho_0>0$ (as in Fig.~\ref{F1}). 

As seen in Fig.~\ref{F1}, where the exact solution is physical, it can serve as a valuable benchmark for assessing the accuracy and efficiency of various hydrodynamic approximation schemes. The power of the Gubser solution lies in the fact that it embodies a key feature of relativistic heavy-ion collisions, namely simultaneous longitudinal and transverse expansion with sharply different expansion rates. In fact, the transverse expansion of the Gubser solution is extremely rapid, with transverse velocity gradients that produce large dissipative flows and therefore severely stress any hydrodynamic framework based on an expansion around local thermal equilibrium, probably more severely than the flow profiles occurring in real heavy-ion collisions. That anisotropic hydrodynamics can capture these large dissipative flows and evolve them with high physical accuracy, as illustrated by the excellent agreement with the exact Gubser solution, suggests that it is a robust framework that yields reliable results even under extreme conditions, such as those existing in the expanding quark-gluon plasma fireballs created in small collision systems such as p+A or p+p.    

\vspace*{-3mm}
\section{Conclusions}
\label{sec5}
\vspace*{-1mm}

Comparing various hydrodynamic approximation schemes with exact solutions of the Boltzmann equation we found the following performance hierarchy:
$$\mathbf{{\sc vaHydro}{\,>\,}{\sc aHydro}{\,>\,}DNMR{\,\sim\,}IS{\,>\,}NS{\,>\,}ideal\,fluid}$$
We expect this to continue to hold for strongly coupled liquids which cannot be described by kinetic theory. 







\end{document}